\documentclass[aps,twocolumn,nofootinbib,showpacs,final,balancelastpage,superscriptaddress]{revtex4}
\usepackage{graphicx}
\usepackage{tabularx}
\usepackage{amssymb}
\usepackage{amsmath}
\usepackage{amsbsy}
\usepackage{comment}
\usepackage{mathrsfs}
\usepackage{dsfont}
\usepackage{nicefrac}
\usepackage{wrapfig}
\usepackage{amsfonts}
\usepackage{url} 
\usepackage[nolist]{acronym}
\usepackage{nicefrac}
\usepackage{subfigure}
\usepackage{epsf,color,colordvi,pifont}
\usepackage{showkeys}

\usepackage{slashed}

\DeclareFontFamily{OT1}{pzc}{}
\DeclareFontShape{OT1}{pzc}{m}{it}{<-> s * [1.10] pzcmi7t}{}
\DeclareMathAlphabet{\mathpzc}{OT1}{pzc}{m}{it}

\begin{document}
\title{Linear Breit-Wheeler pair production by high-energy bremsstrahlung photons colliding with an intense X-ray laser pulse}
\author{A. \surname{Golub}}
\email{Alina.Golub@uni-duesseldorf.de }
\affiliation{Institut f\"{u}r Theoretische Physik I, Heinrich-Heine-Universit\"{a}t D\"{u}sseldorf, Universit\"{a}tsstra\ss{e} 1, 40225 D\"{u}sseldorf, Germany.}
\author{S. \surname{Villalba-Ch\'avez}}
\email{selym@tp1.uni-duesseldorf.de}
\affiliation{Institut f\"{u}r Theoretische Physik I, Heinrich-Heine-Universit\"{a}t D\"{u}sseldorf, Universit\"{a}tsstra\ss{e} 1, 40225 D\"{u}sseldorf, Germany.}
\author{H. \surname{Ruhl}}
\email{hartmut.ruhl@physik.uni-muenchen.de}
\affiliation{Arnold Sommerfeld Center, LMU Munich, Theresienstra{\ss}e 37, 80333 M\"{u}nchen, Germany.}
\author{C. \surname{M\"{u}ller}}
\email{c.mueller@tp1.uni-duesseldorf.de}
\affiliation{Institut f\"{u}r Theoretische Physik I, Heinrich-Heine-Universit\"{a}t D\"{u}sseldorf, Universit\"{a}tsstra\ss{e} 1, 40225 D\"{u}sseldorf, Germany.}

\begin{abstract}
A possible setup for the experimental verification of linear Breit-Wheeler pair creation of electrons and positrons in photon-photon collisions is studied theoretically. It combines highly energetic bremsstrahlung photons, which are assumed to be generated by an incident beam of GeV electrons penetrating through a high-$Z$ target, with keV photons from an X-ray laser field, which is described as a focused Gaussian pulse. We discuss the dependencies of the pair yields on the incident electron energy, target thickness, laser parameters, and collision geometry. It is shown that, for suitable conditions which are nowadays in reach at X-ray laser facilities, the resulting number of created particles seems to be well accessible for enabling the first experimental observation of the linear Breit-Wheeler process $\gamma\gamma\to e^+e^-$.
\end{abstract}


\date{\today}

\maketitle

%%%%%%%%%%%%%%%%%%%%%%%%%%%%%%%%%%%%%%%%%%%%%%%%%%%%%%%%%%%%%%%%%%%%%%%%%%%%%%%%%%%%%%%%%%%%%%%%%%%%%%%%%%%%%%%%%%%%%%%%%%%%%%%%%%%%
\begin{figure*}[ht]
\includegraphics[width=\textwidth]{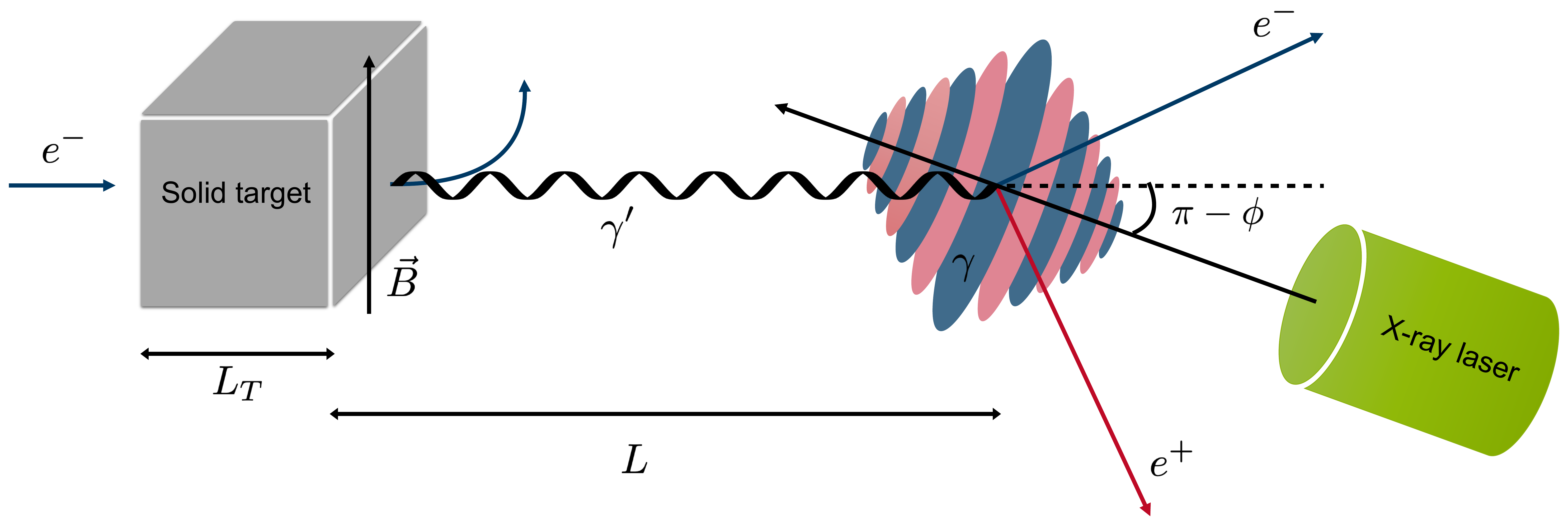}
\caption{\label{fig1} Setup of the proposed experiment for electron ($e^-$) and positron ($e^+$) creation. Here, the GeV and keV photons involved in the reaction are provided by bremsstrahlung and X-ray laser radiation, respectively.}
\end{figure*}

\section{Introduction}
Production of electrons and positrons in a collision of photons represents one of the most intriguing predictions of quantum electrodynamics (QED) as it implies that matter can be created from light. Although the underlying fundamental process $\gamma\gamma\to e^+e^-$ was described theoretically by Breit and Wheeler long ago \cite{BreitWheeler}, experimental validation of electron-positron pair production from light has so far been accomplished solely in the nonlinear multi-photon regime, as part of a two-step process \cite{Burke, Bamber}: firstly, the interaction of a 46.6 GeV electron beam with an optical terawatt laser pulse at the Stanford Linear Accelerator Center (SLAC) provided backscattered photons of multi-GeV energy which, subsequently, created electron-positron pairs upon collision with several eV laser photons (nonlinear Breit-Wheeler process \cite{RitusReview}). Following energy-momentum conservation with inclusion of field-dressing effects, at least five laser photons were needed in the second reaction step to overcome the energy threshold \cite{Reiss2009, Hu2010}, and a total of about 100 positrons was detected during the experiment.

The successful SLAC experiment, in combination with the ongoing development of high-intensity laser technology, has triggered a substantial theoretical interest in laser-induced pair production and, particularly, in the Breit-Wheeler process \cite{EhlotzkyReview, diPiazzaReview}. On the one side, there is a research focus on the highly nonlinear regime of this process \cite{Krajewska2012, Krajewska2014, Selym, Jansen2013,  Meuren, Blackbourn, diPiazza}, where its rate exhibits an exponential nonperturbative field dependence resembling the famous Schwinger rate. 
  Besides, the influence of the precise shape of the laser field has been examined in detail (see \cite{diPiazza, Heinzl, Kaempfer2012, Kaempfer2016, Jansen2016, Jansen2017, Grobe, Kaempfer2018} and references therein). On the other side, also the linear version of the process \cite{CommentLinear}, as originally studied by Breit and Wheeler, where an electron-positron pair is produced in a collision of two photons, has been under active scrutiny by theoreticians \cite{KingGies, Pike, Drebot, Ribeyre, Yu, Golub}. While being known to play a crucial role in various astrophysical contexts, such as $\gamma$-ray bursts, black hole dynamics and active galactic nuclei \cite{Piran, Ruffini}, this elementary QED effect has not been validated experimentally in the laboratory yet.

The difficulty for observing the linear Breit-Wheeler effect manifests itself when the energy-momentum conservation is considered 
\begin{equation}
k + k' = p_++p_-,
\end{equation}
where $k$ and $k'$ stand for the photons wave vectors and $p_{\pm}$ denote the electron and positron four-momenta, correspondingly. It leads to the threshold relation $\omega \omega' \geq \frac{2m^2}{1-\mathrm{cos}(\phi)}$ between the photon energies $\omega$ and $\omega'$, the particles' rest mass $m$, and the angle $\phi$ between the photon propagation directions [see Fig.~\ref{fig1}]. Accordingly, in the center-of-mass frame, the photon energies need to be of the MeV order---a scenario, which is challenging to achieve nowadays with sufficiently high beam intensities. Throughout this paper we work in Lorentz-Heaviside units where $c=\hbar=\epsilon_0=1$ and use the metric with signature $\mathrm{diag}(g^{\mu \nu})=(1,-1,-1,-1)$. 

In recent years, various theoretical proposals for detection of linear Breit-Wheeler pair production were put forward. The first designs relied on photon-photon colliders, where one source was provided by thermal hohlraum radiation, whereas the partner photon was either described as a plane electromagnetic wave \cite{KingGies} or supposed to result from bremsstrahlung inside a high-$Z$ target \cite{Pike}. These proposals were followed by symmetric setups incorporating photons stemming from two equal origins, such as Compton gamma sources \cite{Drebot}, laser pulses interacting with thin aluminium targets or dense, short gas jets \cite{Ribeyre} or 10-petawatt laser beams penetrating through narrow tube targets \cite{Yu}. Additionally, the possibility of detecting an analog of the Breit-Wheeler process in bandgapped graphene layers at much lower energy scale has recently been discussed \cite{Golub}.

In the current paper, we propose an alternative approach to detect the linear Breit-Wheeler process, which seems to be feasible in the nearest future, as the involved techniques are well established nowadays. The setup under consideration relies on few-GeV bremsstrahlung photons interacting with keV photons from an X-ray free-electron laser (XFEL) beam, this way providing the energy needed for the process to take place. The bremsstrahlung is assumed to be generated by a GeV electron beam penetrating through a high-$Z$ solid target; as a possible source of the highly relativistic incident electrons we propose laser wakefield acceleration (LWFA), which has proven itself to be suitable for the production of narrowly collimated intense electron beams \cite{Leemanns2019,Karsch}. 
We point out that the required ingredients for our scheme are in principle available at the HiBEF facility \cite{HiBEF} at DESY (Hamburg, Germany), where a short-pulse 300-terawatt optical laser is operated in conjunction with the European XFEL.   

It is important to note that Bethe-Heitler pair production in the high-$Z$ target represents a competing channel for positron creation, as the bremsstrahlung photons may decay into pairs in the atomic Coulomb fields of the solid \cite{Chen, Sarri}. Therefore, in order to ensure the interaction volume being free from disturbance caused by electrons from the incident beam or potentially created Bethe-Heitler pairs, a static magnetic field needs to be introduced sidetracking the charged particles. Other QED processes are neglectable in the considered parameter range. A schematic diagram of the proposed setup is shown in Fig.~\ref{fig1}. We point out that a similar setup was discussed in the context of nonlinear, nonperturbative Breit-Wheeler pair creation, where---instead of an intense X-ray laser---an optical petawatt laser was involved \cite{Blackbourn}.

Our paper is structured in the following way: after the introduction, we present in Sec.~II the theoretical framework for describing the linear Breit-Wheeler process in a collision of a highly energetic bremsstrahlung photon with an X-ray laser photon. The laser light is modeled as a linearly polarised, monochromatic Gaussian pulse and the spectrum of bremsstrahlung radiation is taken into account. Section III illustrates our results and provides estimations for the number of created positrons which can be detected in dependence on the various setup parameters. In the last section a conclusion is provided and, finally, some technical details are given in an appendix.

%%%%%%%%%%%%%%%%%%%%%%%%%%%%%%%%%%%%%%%%%%%%%%%%%%%%%%%%%%%%%%%%%%%%%%%%%%%%%%%%%%%%%%%%%%%%%%%%%%%

\section{Theoretical description} \label{sec:QKA}
This section is devoted to a theoretical perturbative development of linear electron-positron pair creation occurring when bremsstrahlung photons impinge on photons coming from a laser field. In contrast to the well-established consideration, where both photons are treated as quantized modes, we model the laser photons as a classical Gaussian pulse. In this context the Volkov-states approach, which is customarily pursued in strong-field QED \cite{RitusReview}, is not applicable as we are dealing with a focused field. Hence, we shall firstly formulate a theoretical framework which--starting from a fully quantum picture--allows us to go smoothly over to a description of the interaction between a classical laser field and an arbitrary quantized photon mode (Sec.~II.~A). This approach will afterwards enable to incorporate the field focussing (Sec.~II.~B) as well as the spectrum of bremsstrahlung photons (Sec.~II.~C) in a natural way.
%%%%%%%%%%%%%%%%%%%%%%%%%%%%%%%%%%%%%%%%%%%%%%%%%%%%%%%%%%%%%%%%%%%%%%%%%%%%%%%%%%%%%%%%%%%%%%%%%%%

\subsection{Pair production in a laser field}
In the parameter range of interest--where the maximum amplitude $\mathcal{E}_0$ of the laser electric field satisfies $e\mathcal{E}_0\ll m\omega$--we describe the linear Breit-Wheeler process by the second-order scattering matrix element $S_{fi}$, where the initial quantized photon and the created electron-positron pair are taken into account by number states $\vert k',\epsilon' \rangle$ and $\vert p_{+},\sigma_+; p_-, \sigma_-\rangle$, correspondingly. Here, $k'$ is the wave four-vector of the involved photon with polarization $\epsilon'$, whereas $p_{\pm}$, $\sigma_{\pm}$ stand for the electron and positron four-momenta and spin states. On the other hand, the description of a laser field may be accomplished by introducing a coherent state $\vert \alpha_k \rangle$ in the mode $k$. 
 Since the coherent description insures the incorporation of large numbers of photons in the laser field, the initial and final coherent states can be approximately considered the same, as only one laser photon is involved in the process. Accordingly, we obtain
\begin{multline}\label{sfi}
S_{fi}=\langle p_{+},\sigma_+; p_-, \sigma_-;\alpha_{k} \vert \hat{S}[\hat{\psi},\hat{\bar{\psi}},\hat{a}]\vert k', \epsilon'; \alpha_{k}\rangle \to \\
 \langle p_{+},\sigma_+; p_-, \sigma_- \vert \hat{S}[\hat{\psi},\hat{\bar{\psi}},\hat{a}+a_{ext}]\vert k',  \epsilon' \rangle
\end{multline}
with the QED scattering operator $ \hat{S}[\hat{\psi},\hat{\bar{\psi}},\hat{a}]$ of the second order in the fermion $\hat{\psi}$ and gauge $\hat{a}$ field operators
\begin{multline}
\hat{S}[\hat{\psi},\hat{\bar{\psi}},\hat{a}] =\\
 \frac{(ie)^2}{2!}\int d^4x \ d^4y \ \hat{\cal{T}}[\hat{\bar{\psi}}(x)\hat{\slashed{a}}(x)\hat{\psi}(x)\hat{\bar{\psi}}(y)\hat{\slashed{a}}(y)\hat{\psi}(y)],
\end{multline}
where we have introduced the Feynman notation $\hat{\slashed{a}}=\gamma^\mu a_\mu$ with Dirac gamma matrices $\gamma^\mu$, $\hat{\bar{\psi}}=\hat{\psi}\gamma^0$ and  $\hat{\cal{T}}$ stands for the time-ordering operator.
In the second line of Eq.~\eqref{sfi} the results obtained in \cite{Fradkin} were used. This procedure allows us to evaluate the consequence of the coherent state by adding to the photon field operator $\hat{a}^\mu(x)$ the classical electromagnetic field potential in Lorenz gauge $a^\mu_{ext}(x)=\epsilon^\mu a_0(x,k)$ with polarization $\epsilon^\mu$ ($\epsilon^\mu k_\mu=0$). The quantization of the former has been carried out within the Gupta-Bleuler formalism.
Based on the expression in the second line of Eq.~\eqref{sfi} we can calculate the Breit-Wheeler process for a classical laser field of arbitrary shape, since the latter may be written as a linear superposition of wave modes. Accordingly, $a_0(x,k)$ shall represent the vector potential amplitude of a focused laser field with central wave four-vector $k^\mu$.

The rate per volume of the process is obtained as
\begin{equation}\label{rate1}
R^{\sigma_\pm}_{\epsilon, \epsilon'} \ (k,k') = \int \frac{V d^3p_+}{(2\pi)^3}\frac{V d^3p_-}{(2\pi)^3}\frac{\vert S_{fi} \vert ^2}{TV},
\end{equation}
where the scattering matrix element is being integrated over the phase space of created electron and positron, while divided by the interaction time $T$ and volume $V$.
In our context, after averaging over the polarization of the quantized photon as well as summation over the fermions spins, the rate per volume reads
\begin{equation}\label{rate2}
R(k,k')  = \frac{1}{TV}\int \frac{d^4\tilde{k}}{(2\pi)^4}\vert \tilde{a}_0(\tilde{k},k)\vert^2\frac{R_{BW}(\tilde{k}k')}{N_{\gamma}^2}
\end{equation}
with the Fourier transform of the vector potential amplitude
\begin{equation}
\tilde{a}_0(\tilde{k},k)=\int d^4x\,\mathrm{e}^{i\tilde{k}x}a_0(x,k).
\end{equation}
Moreover, the quantity $R_{BW}(\tilde{k}k')/N_{\gamma}^2$ stands for the well known rate of Breit-Wheeler pair creation obtained as a result of the collision of two gamma quanta
\begin{equation}\label{rateBW}
\begin{split}
& R_{BW}(\tilde{k}k')= \frac{\alpha e^2N_{\gamma}^2}{\omega' V_{\gamma}} u(s), \\
& u(s)=\left[\frac{-s \sqrt{ s^2-1}(1 + s^2)}{s^4} \right.\\
& \qquad \qquad \left. + \frac{(-1 + 2 (s^2 + s^4)) \mathrm{ln}(s+\sqrt{ s^2-1})}{s^4}\right],
\end{split}
\end{equation}
where $N_{\gamma}$ refers to the normalization constant of the quantized field. In this formula,
$\alpha \approx 1/137$ is the fine-structure constant, $V_{\gamma}$ is the quantization volume and $s^2 = kk'/2m^2$ is the normalised Mandelstam variable \cite{Greiner, RitusReview}. 

%%%%%%%%%%%%%%%%%%%%%%%%%%%%%%%%%%%%%%%%%%%%%%%%%%%%%%%%%%%%%%%%%%%%%%%%%%%%%%%%%%%%%%%%%%%%%%%%%%%
\subsection{Pair production in the field of a Gaussian pulse}
In this section we model the laser field by incorporating the vector potential of a linearly polarized Gaussian pulse in paraxial approximation propagating in $z-$direction for $a^{\mu}_{ext}$. As can be seen in Eq.~\eqref{rate2}, for proceeding further, we require the absolute value squared of the Fourier transformed amplitude $a_0(x,k)$, which is given in Appendix \ref{App2}. When considering Eq.~\eqref{a0} the rate per volume in cylindrical coordinates reads
\begin{multline}\label{rate3a}
R(k,k') = \frac{\mathcal{E}_0^2 N_\gamma^2}{32}\frac{(\tau/2)^2 w_0^4}{TA}\int_0^{2\pi}d\phi_{\tilde{k}}\int_{-\infty}^\infty d \tilde{k}_{z} \\
\int_0^{\vert \tilde{k}_0\vert}d\tilde{k}_{\perp} \ \tilde{k}_{\perp} \mathrm{e}^{-\frac{\tilde{k}_{\perp}^2w_0^2}{2}} \int_{-\infty}^\infty \frac{d\tilde{k}_0}{\tilde{k}_0^2} \mathrm{e}^{-\frac{(\tau/2)^2}{2}(\omega-\tilde{k}_0)^2}\\
\delta(\tilde{k}_z-\tilde{k}_0+\frac{\tilde{k}_\perp^2}{2\omega}) R_{BW}(\tilde{k}k').
\end{multline}
Here, $A$ denotes the interaction area resulting from the interaction volume divided by a length factor which stems from squaring the Dirac $\delta$-function. Next, we perform the integration over $\tilde{k}_z$ by exploiting the latter one and, afterwards, integrate over $\tilde{k}_0$ by evaluating all components except the exponential function at $\tilde{k}_0 = \omega$ as it provides the biggest contribution to the integral. In that case, the argument of $R_{BW}$ can be approached by $\tilde{k}k' \approx \omega \omega'\left(1-\mathrm{cos}(\phi_{\tilde{k}}) \sqrt{1+\frac{\tilde{k}_{\perp}^4}{4 \omega^4}}\right)$ and when substituting $v = \frac{\tilde{k}_{\perp} w_0}{\sqrt{2}}$ we obtain
\begin{multline}\label{rate3}
R(k,k') \approx \frac{\mathcal{E}_0^2N_\gamma^2}{4\omega^2}\frac{(\tau/2)}{TA}\sqrt{\frac{\pi}{2}}\frac{w_0^2}{2}
\int_0^{\frac{\omega w_0}{2}}dv \ v \mathrm{e}^{-v^2} \\
\times \int_0^{2\pi}d\phi_{\tilde{k}} R_{BW}\left[ \omega \omega'\left(1-\mathrm{cos}(\phi) \sqrt{1+\frac{v^4}{w_0^4 \omega^4}}\right) \right],
\end{multline}
The upper bound of integration in the radial component is restricted to $\frac{\omega w_0}{2}$ as $\tilde{k}_{\perp}$ is bounded to lie within the interval $[0,\omega]$. Moreover, we treat the collision angle $\phi$ between laser and quantized photon as an external parameter which is predefined in our setup [see Fig.~\ref{fig1}]. Its value will be chosen in a way to allow for a practicable geometry of the experimental setup which avoids damaging of technical devices by the intense laser beam.

When taking into account the paraxial approximation ($\tilde{k}_{\perp} \ll \omega$), the square root in $R_{BW}$ can be approached by 1 and, consequently, the rate of the process becomes
\begin{multline}\label{ratePhi}
R(\omega, \omega', \mathrm{cos}(\phi))\approx \\
\frac{a_0^2}{4N_\gamma^2}
(1-\mathrm{e}^{-\frac{w_0^2\omega^2}{4}})
 R_{BW}\left[ \omega \omega'(1-\mathrm{cos}(\phi)) \right].
\end{multline}
Here, we have used the relations $TA=\frac{\tau}{2}\sqrt{\frac{\pi}{2}}\frac{\pi w_0^2}{2}$ as the laser pulse energy of the considered field reads $E_L=\frac{\mathcal{E}_0^2}{2}\frac{\tau}{2}\sqrt{\frac{\pi}{2}}\frac{\pi w_0^2}{2}$ \cite{Blinne}. Additionally, we denote $a_0=\mathcal{E}_0/\omega$.
%%%%%%%%%%%%%%%%%%%%%%%%%%%%%%%%%%%%%%%%%%%%%%%%%%%%%%%%%%%%%%%%%%%%%%%%%%%%%%%%%%%%%%%%%%%%%%%%%%%%%%%%%%%%%%%%%%%%%%%%%%%%%%%%%%%%%%%%%%%%%

\subsection{Bremsstrahlung photons}
To obtain the pair creation rate resulting from a collision of bremsstrahlung photons with a Gaussian laser pulse we integrate the rate, as given in Eq.~\eqref{rate3}, weighted by the distribution function of the bremsstrahlung photons $W(k')$ with respect to the photon momentum $k'$:
\begin{equation}
R_\gamma(k)= \int \frac{d^3 k'}{(2\pi)^3}W(k')R(k,k').
\end{equation}
Since the incoming electrons employed in the proposed setup are highly relativistic, the generated bremsstrahlung photons will be emitted preferably in the direction of the incident electron propagation. More precisely, the angle of photon spreading can be approximated by $\theta \approx \frac{m\mathrm{ln}(E_0/m)}{E_0}$ \cite{Stearns}, where $E_0$ stands for the initial kinetic energy of the incoming electrons. Hence, for $E_0$ of the order of GeV we will have angles in the mrad range. Considering this fact allows us to approximate the bremsstrahlung distribution function in spherical coordinates by 
 \begin{equation}\label{thin}
W(k') \approx \frac{(2\pi)^3I_\gamma(f,\ell)}{\omega'^2\mathrm{sin}(\theta_{k'})E_0} \Theta(E_0-\omega')
\delta(\theta_{k'}-\phi)\delta(\phi_{k'})
\end{equation} 
with the photon energy spectrum derived within the complete screening approximation \cite{Tsai}
\begin{equation}\label{thick}
I_\gamma(f,\ell) \approx \frac{(1-f)^{\frac{4}{3}\ell}-\mathrm{e}^{-\frac{7}{9}\ell}}{f(\frac{7}{9} +\frac{4}{3} \mathrm{ln}(1-f))}
\end{equation}
and $\phi$ being the angle between the directions of propagation of laser photons and bremsstrahlung electrons as both beams lie in one plane, whereas the laser beam propagates in $z$ direction and its focal point is set to define the origin [see Eq.~\eqref{EParApp}]. Moreover,  the distribution function depends on the normalised target thickness $\ell=L_{T}/L_{rad}$ with $L_{rad}$ being the radiation length of the target material and $f = \omega'/E_0$ stands for the normalised photon energy. Notice that the formula above provides a good estimation for the photon energy distribution in the ranges $0.5\lesssim \ell \lesssim 2$ and $f\gtrsim 0.2=f_{\rm min}$ (see \cite{Tsai}). 

Hence, when taking into account the assumptions listed above Eq.~\eqref{ratePhi}, the rate per volume of the process reads
\begin{equation}\label{ratebremsstr}
R_{\gamma}(\omega,\mathrm{cos}(\phi))=\int_0^1df R(\omega, \omega', \mathrm{cos}(\phi)) \ I_{\gamma}(f,\ell).
\end{equation}
This equation constitutes the basis for our numerical results which are presented below.

%%%%%%%%%%%%%%%%%%%%%%%%%%%%%%%%%%%%%%%%%%%%%%%%%%%%%%%%%%%%%%%%%%%%%%%%%%%%%%%%%%%%%%%%%%%%%%%%%%%%%%%%%%%%%%%%
\section{Results and Discussion}
\begin{figure*}[ht]
\includegraphics[width=0.4\textwidth]{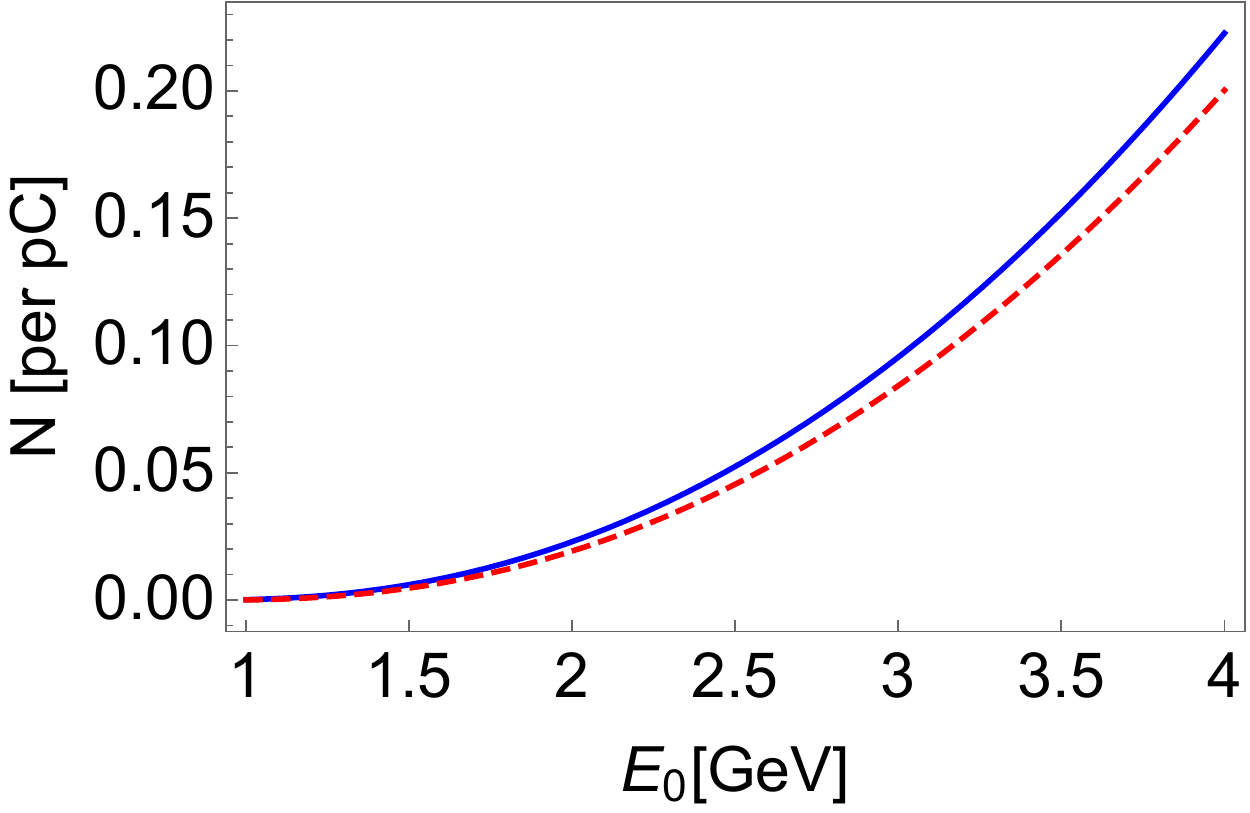}
\includegraphics[width=0.4\textwidth]{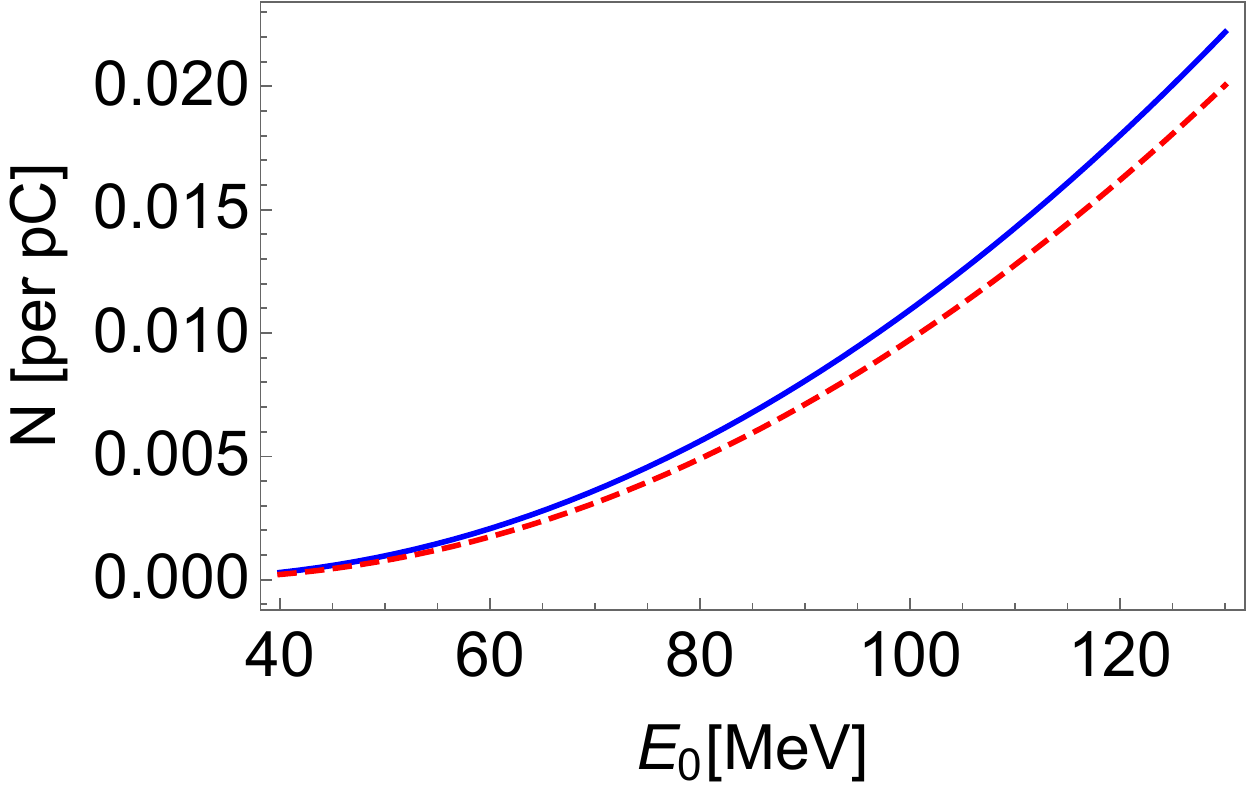}
\caption{\label{fig2} Red ($\phi = 150^{\circ}$) and blue ($\phi = 165^{\circ}$) curves show the number of pairs in dependence on the initial electron energy for $\xi=0.001$, $\tau= 100 \ \mathrm{fs}$, $w_0 =20 \ \mu\mathrm{m}$, $L=50 \ \mathrm{cm}$ and $L_T=7 \ \mathrm{mm}$. Laser frequencies are set to $ \omega = 0.3\ \mathrm{keV} $ for the left panel and $ \omega = 10\ \mathrm{keV} $ for the right panel.}
\end{figure*}

When aiming to provide an estimation of the average number of created electron-positron pairs per incident electron emitting bremsstrahlung, we multiply the rate of the process as given in Eq.~\eqref{ratebremsstr} with the laser pulse duration and interaction volume
\begin{equation}
\mathcal{N}=\tau V R_\gamma.
\end{equation} 
In the process under consideration, there is an interaction volume provided by the laser pulse which can be approximated by $V=w_0^2\pi\tau$. Moreover, the spreading beam of bremsstrahlung photons introduces a second constraint on the area, where the desirable reaction is possible. In our framework this coincides with the photon quantization volume $V_{\gamma}$ as given in Eq.~\eqref{rateBW}. With the small-angle approximation it can be estimated by $\pi(\theta L)^2\tau$, where $\theta$ is the photon collimation angle and $L$ stands for the distance between the solid target and laser focus [see Fig.~\ref{fig1}].  Hence, the number of pairs per incident electron reads
\begin{equation}\label{N}
\mathcal{N} \approx \frac{\tau w_0^2}{(\theta L)^2} \frac{\alpha \xi^2m^2}{4E_0}\int_{f_{\rm min}}^1 \frac{df}{f} I_{\gamma} (f,l)u(s)
\end{equation}
as for an X-ray laser beam the wave length is much smaller than the beam waist $w_0\gg 1/ \omega$ and we perform the integration in the region where Eq.~\eqref{thick} provides a good approximation. Here, the normalised Mandelstam variable has been expressed as $s=\sqrt{\frac{\omega E_0f(1-\mathrm{cos}(\phi))}{2m^2}}$ [compare with Eq.~\eqref{rateBW}]. Moreover, for the following discussion we introduce the usual laser field strength parameter $\xi=ea_0/m$, which serves as an indicator of whether the perturbative procedure can be applied. In our context we restrict the parameter space of $\xi$ to lie well below 1. 

Next, let us examine how the number of pairs, as given in Eq.~\eqref{N}, depends on the setup parameters. The rapidly evolving field of laser-wakefield electron acceleration allows for compact experimental arrangements providing electrons with energies of up to several GeV when using a subpetawatt-class laser \cite{Leemans2014}, as provided, for example, within the HiBEF project at the European XFEL \cite{HiBEF}. Here, we will cover electron energies in the range of 40 MeV to 4 GeV. 
By penetrating through a lead target, these electrons generate a spectrum of bremsstrahlung with corresponding endpoint energies.
As the Eq.~\eqref{ratebremsstr} represents a reliable approximation to the bremsstrahlung spectrum for restricted values of $f$, we 
will chose suitable laser frequencies in the domain from soft to hard X-rays ($0.3-10$ keV) in order to meet the conditions of applicability. Soft X-ray laser pulses of 0.3 keV photon energy could be delivered, for example, by the FLASH facility at DESY in Hamburg, where photons with wavelength between $4.2-52$~nm can be generated. Hard X-ray laser pulses with photon energies of 10 keV and even higher are available at the European XFEL at DESY and the LCLS at Stanford \cite{LCLS}. For our numerical calculations,  we choose the value of the laser field strength parameter  as $\xi=0.001$ throughout. It corresponds to an intensity of $I \approx8\times 10^{16} \ \mathrm{W/cm^2}$ at $\omega=0.3$ keV and $I \approx9\times 10^{19} \ \mathrm{W/cm^2}$ at $\omega=10$ keV.

Figure~\ref{fig2} shows how the number of created positrons, which in our case coincides with the number of created pairs, depends on the energy $E_0$ of the incident electron beam. The left panel considers the collision of a soft X-ray laser pulse ($\omega = 300$ eV) with bremsstrahlung emitted from $1-4$ GeV electrons. The right panel assumes, instead, that the pairs are created by an XFEL pulse of 10 keV photon energy which collides with bremsstrahlung from $40-130$ MeV incident electrons.
The red ($\phi = 150^{\circ}$) and blue ($\phi = 165^{\circ}$) curves correspond to different angles between the colliding photons [see Fig.~\ref{fig1}]. We see that, in the considered energy ranges, the number of pairs grows with increasing $E_0$. Even though the center-of-mass energy available for pair creation is similar in the left and right panels, respectively, the number of produced pairs is about ten times larger on the left. This outcome can be attributed to the fact that, for smaller $E_0$, the bremsstrahlung emission angle $\theta$ increases, which reduces the overlap with the focal region of the laser pulse and, thus, the number of bremsstrahlung photons in the interaction volume. In comparison with the aforementioned effect, the impact of the prefactor $1/\omega'$ in Eq.~\eqref{rateBW} is of minor importance here. We note that, for the largest electron energy considered ($E_0=4$ GeV) slightly more than 2 pairs can be generated per 10 pico-Coulomb of incident electrons.

\begin{figure}[ht]
\includegraphics[width=0.4\textwidth]{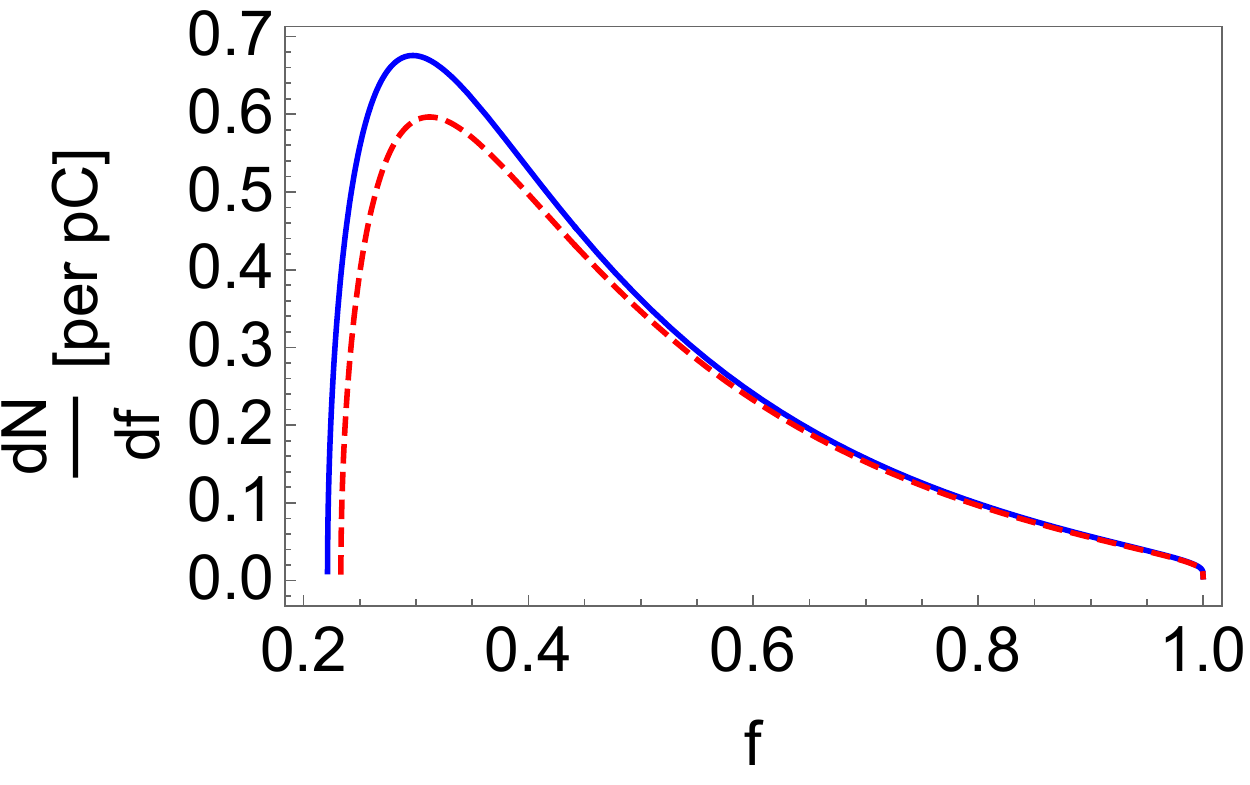}
\caption{\label{fig2a} Contribution of bremsstrahlung frequencies to the pair production for a lead target of thickness $L_T= 7 \ \mathrm{mm}$ and $E_0 = 4\ \mathrm{GeV}$, $\xi=0.001$, $ \omega = 0.3\ \mathrm{keV} $, $\tau= 100 \ \mathrm{fs}$, $w_0 =20 \ \mu\mathrm{m}$, $L=50 \ \mathrm{cm}$ for $\phi = 165^{\circ}$ (blue) and $\phi = 150^{\circ}$ (red).}
\end{figure}

For an initial electron energy of $E_0 = 4$ GeV we examine the functional dependence of the integrand in Eq.~\eqref{N} on the normalized bremsstrahlung photon energy $f$ in Fig.~\ref{fig2a}. The course of the curves reflects several tendencies. While the factor u(s) in Eq.~\eqref{rateBW} always increases with growing photon energy, the Breit-Wheeler rate $R_{BW}$ itself first quickly grows from zero at the threshold, reaches a maximum at a particular
value of $\omega'$, and afterwards starts to decay--this way reflecting
the influence of the factor $1/\omega'$ in Eq.~\eqref{rateBW}. For our parameter
set the value of $\omega'$ leading to the maximum rate is approximately $1.5$ GeV. The latter tendency is enhanced by the fact that the number of bremsstrahlung photons falls when their energy rises [as provided by Eq.~\eqref{thick}]. Due to the combination of these effects, the maximum contributions to the pair yields stem from the spectral region around $\omega' \approx 0.3E_0$. Accordingly, the electrons and positrons are created with typical energies of about $600$ MeV. They are emitted predominantly in the propagation direction of the bremsstrahlung photons. In addition, we see in Fig.~\ref{fig2a} that the maximum of the red curve for $\phi = 150^{\circ}$ is slightly shifted to the right. This effect is caused by the $\phi$-dependence of the threshold energy $\sim[1-\cos(\phi)]^{-1}$. The smaller $\phi$, the larger $\omega'$ must be to overcome it.

\begin{figure}[ht]
\includegraphics[width=0.4\textwidth]{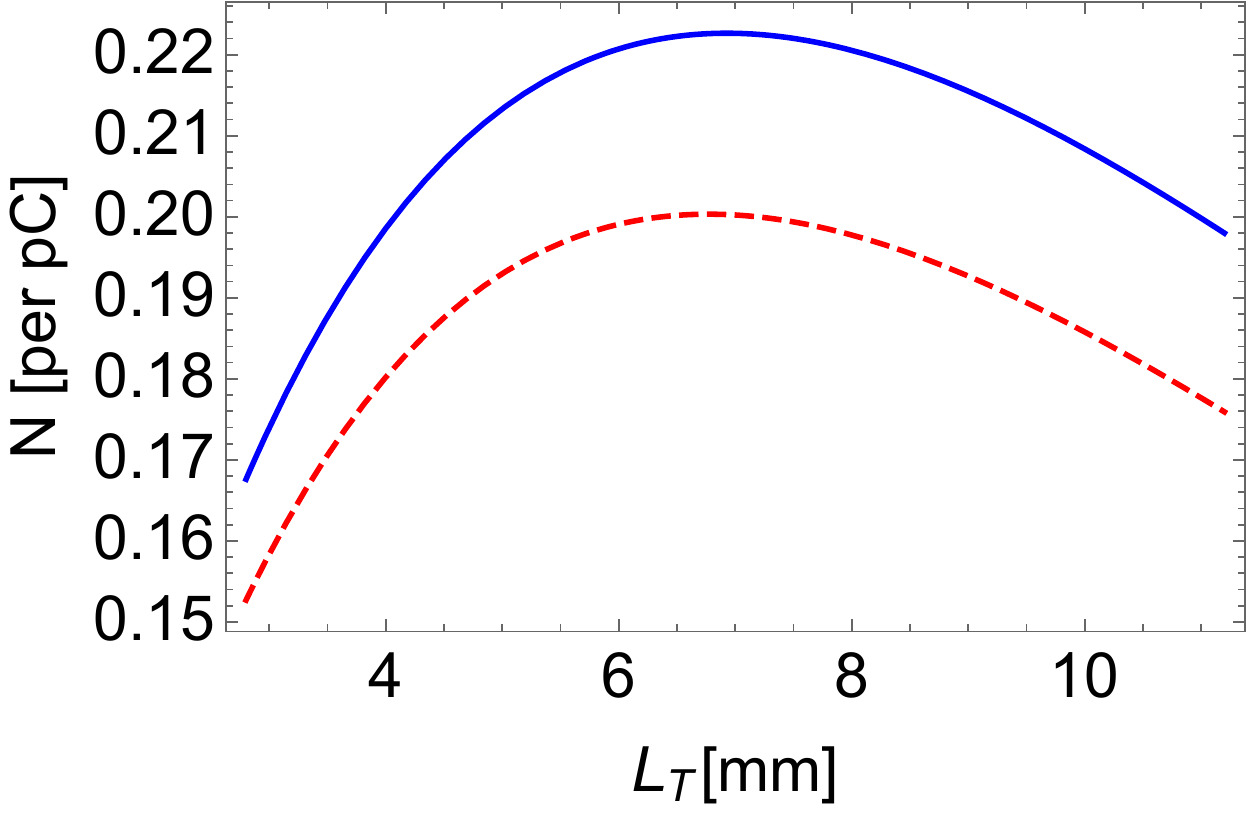}
\caption{\label{fig3} Number of created pairs per pC for different target thicknesses $L_T$ of lead and $E_0 = 4\ \mathrm{GeV}$, $\xi=0.001$, $ \omega = 0.3\ \mathrm{keV} $, $\tau= 100 \ \mathrm{fs}$, $w_0 =20 \ \mu\mathrm{m}$, $L=50 \ \mathrm{cm}$ for $\phi = 165^{\circ}$ (blue) and $\phi = 150^{\circ}$ (red).}
\end{figure}

Further, Fig.~\ref{fig3} depicts the relation between the thickness of the chosen lead target ($Z=82$, $L_{rad} = 5.6 \ \mathrm{mm}$) and the expected number of created positrons. Both curves firstly grow with increasing target thickness, until they reach their maximum at approximately 7 mm, from where on they decline. The maximum arises because, on the one hand, the probability for the emission of bremsstrahlung grows when the incident electrons have to travel through the target material over longer distances. On the other hand, however, emmitted bremsstrahlung photons can be scattered or reabsorbed in the target; the corresponding probability increases with the target thickness as well. If the latter exceeds a certain value, the photon loss processes start to dominate over their generation, this way causing the appearance of an optimal thickness.

\begin{figure}[ht]
\includegraphics[width=0.4\textwidth]{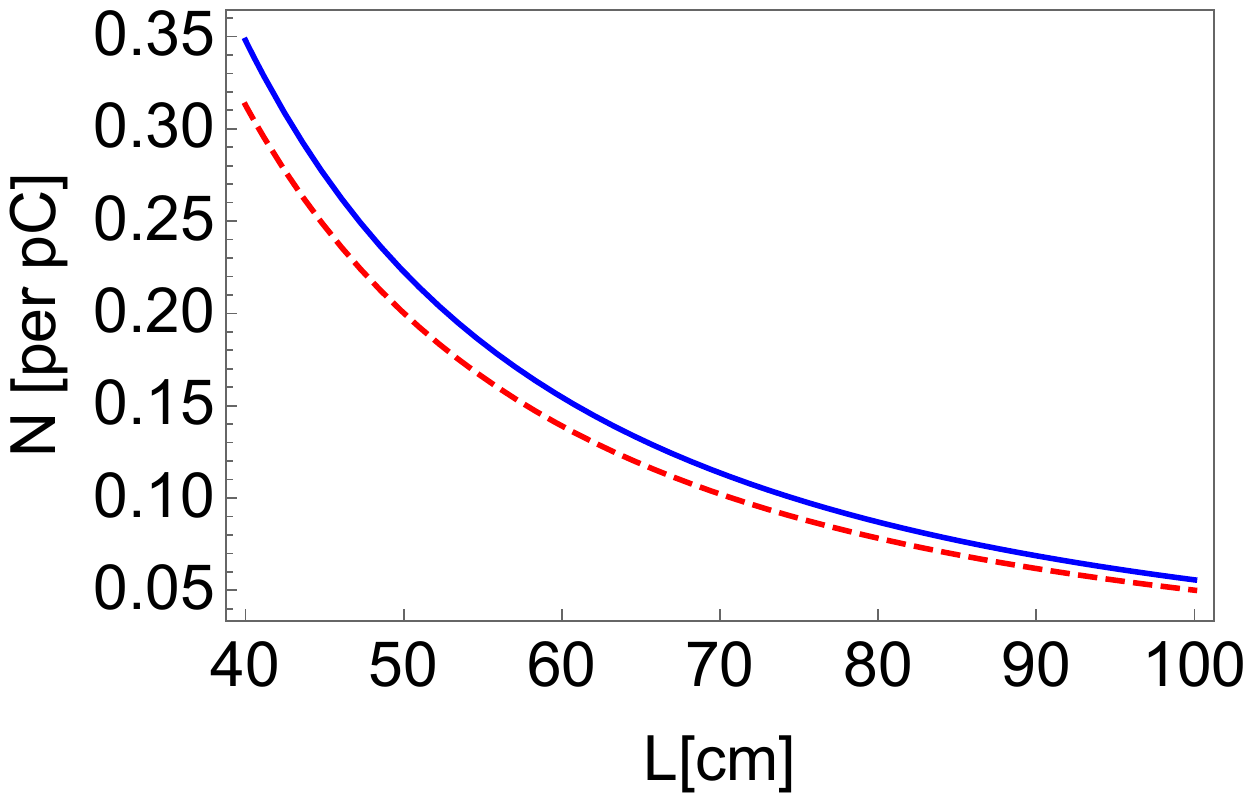}
\caption{\label{fig4} Dependence of the number of created pairs per pC on the distance $L$ between solid target and laser focus. Here, the blue and red curves represent the collision angles $\phi = 165^{\circ}$ and $\phi = 150^{\circ}$, while the remaining  parameters are set to $E_0 = 4\ \mathrm{GeV}$, $\xi=0.001$, $ \omega = 0.3\ \mathrm{keV} $, $\tau= 100 \ \mathrm{fs}$, $w_0 =20 \ \mu\mathrm{m}$ and $L_T=7 \ \mathrm{mm}$.}
\end{figure}

Finally, Fig.~\ref{fig4} illustrates how many pairs can be observed for the optimal parameter set ($E_0 = 4\ \mathrm{GeV}$, $L_T=7 \ \mathrm{mm}$) when changing the distance between X-ray laser focus and solid target. We see that the number of pairs to be detected decreases with growing $L$. The number scales with $L^{-2}$ as Eq.~\eqref{N} shows. Similarly to changes in the photon spreading angle $\theta$, we reduce the number of interacting photons when increasing the distance between the photon sources. It is interesting to note that tighter focusing of the X-ray laser does not provide a similar effect: as the number of created pairs in Eq.~\eqref{N} is quadratic in both the beam waist $w_0$ and the parameter $\xi$, with the latter implying a linear dependence on the laser intensity, the pair yield in the linear Breit-Wheeler regime depends only on the total laser energy $E_L$.

Since the production of incident electron bunches of several 100 pC up to $\approx 1$ nC is becoming feasible with LWFA \cite{Leemanns2019,Karsch}, a detection of up to $\approx 350$ Breit-Wheeler pairs per shot appears approachable with experimental parameters available today or in the near future. This amount of pairs is larger than the total positron yield detected in the SLAC experiment \cite{Burke, Bamber}. It can be compared to predictions provided by similar setups involving highly energetic bremsstrahlung photons for linear and nonlinear Breit-Wheeler processes. In the linear regime of Ref.~\cite{Pike}, up to $10^5$ pairs were obtained for $10^9$ incident electrons, corresponding to approximately 150 pC; the larger pair yield, in comparison with the present setup, may be attributed to the larger interaction volume provided by the radiation-filled hohlraum. Up to $10^4$ pairs per pC were found in the highly nonlinear regime ($\xi=30$) of Ref.~\cite{Blackbourn}. This predicted yield assumes that the detrimental impact of the bremsstrahlung beam divergence can be compensated by suitable prefocusing of the incident electron beam. Otherwise, the number of created pairs would be reduced by several orders of magnitude, accordingly, and would reach a similar level like in the present study.

%%%%%%%%%%%%%%%%%%%%%%%%%%%%%%%%%%%%%%%%%%%%%%%%%%%%%%%%%%%%%%%%%%%%%%%%%%%%%%%%%%%%%%%%%%%%%%%%%%%%%%%%%%%%%%%%
\section{Conclusion and outlook\label{conclusions}}
In the present paper we have theoretically examined a possible setup for detection of linear Breit-Wheeler pair creation, resulting from the interaction of an intense X-ray laser pulse and a beam of high-energy bremsstrahlung. We have shown that by using few-GeV bremsstrahlung photons--produced from several 100 pC of laser-accelarated incident electrons penetrating a few-mm thin high-$Z$ target--and a 100\,fs soft XFEL pulse with $\sim 10^{17} \ \mathrm{W/cm^2}$ intensity, which is feasible today in the laboratory, on the order of $10^2$--$10^3$ 
positrons are expected to be created per shot. The proposed scheme thus offers a way to accomplish the long overdue experimental verification of this fundamental QED process in the linear regime, which was predicted more than 80 years ago.

%%%%%%%%%%%%%%%%%%%%%%%%%%%%%%%%%%%%%%%%%%%%%%%%%%%%%%%%%%%%%%%%%%%%%%%%%%%%%%%%%%%%%%%%%%%%%%%%%%%%%%%%%%%%%%%%%%%%%%%%%%%%%%%%%

\begin{acknowledgments}
This work has been funded by the Deutsche Forschungsgemeinschaft (DFG) under Grant No. 416699545 
within the Research Unit FOR 2783/1. We thank R. W. A. Janjua for his contribution at an early stage of this study and L. Reichwein for useful discussions
on LWFA.
\end{acknowledgments}

\appendix
\section{Fourier transform of a Gaussian pulse's vector potential}\label{App2}
Let us consider the electric field component of a Gaussian pulse
 $\pmb{\mathcal{E}}=\pmb{\epsilon}\mathcal{E}(x,k)$ with linear polarization $\pmb{\epsilon}$ and propagation direction $z$. Within the paraxial approximation $\mathcal{E}(x,k)$ reads
\begin{equation}\label{EParApp}
\mathcal{E}(x,k)=\mathcal{E}_0\frac{\mathrm{e}^{-\frac{(z-t)^2}{(\tau/2)^2}}}{\sqrt{1+\zeta(z)^2}}\mathrm{e}^{-\frac{r^2}{w^2(z)}}\mathrm{cos}(\Phi(x,\omega)),
\end{equation}
where $\mathcal{E}_0$ is the field amplitude, $\tau$ is the pulse duration, $r^2 = x^2+y^2$ and $w(z)=w_0\sqrt{1+\zeta(z)^2}$. Here, $w_0$ is the  waist size of the beam, whereas $\zeta(z)=z/z_R$ with $z_R=w_0^2\pi/\lambda$ denoting the Rayleigh length. Moreover, in the expression above the phase is given by 
\begin{equation}
\Phi(x,\omega)=\omega(z-t)+\zeta(z)\frac{r^2}{w^2(z)}-\mathrm{arctan}(\zeta).
\end{equation}
The field in \eqref{EParApp} can be related to a vector potential $a^\mu = (\varphi, \pmb{\epsilon}a_0)$ fulfilling the Lorenz gauge condition with $\varphi=0$. Indeed, under such restrictions $\pmb{\mathcal{E}}=-\pmb{\epsilon}\frac{\partial a_0}{\partial t}$ holds. Hence, the Fourier transform of the vector potential amplitude results in
\begin{equation}
\tilde{a}_0(\tilde{k},k)=-\int d^4x \mathrm{e}^{i\tilde{k}x}\int^tdt'\mathcal{E}((t',\pmb{x}),k).
\end{equation}

For the sake of convenience, we firstly integrate over the space coordinates and subsequently perform the integrals over $t'$ and $t$. Consequently, we obtain
\begin{multline}
\tilde{a}_0(\tilde{k},k)=-\frac{i\pi^2\sqrt{\pi}w_0^2\mathcal{E}_0}{\tilde{k}_0}\frac{\tau}{2}\mathrm{e}^{-\frac{\tilde{k}_{\perp}^2w_0^2}{4}} \times \\
\left[ \mathrm{e}^{-\frac{(\tau/2)^2}{4}(\omega-\tilde{k}_0)^2}\delta\left(\tilde{k}_z-\tilde{k}_0+\frac{\tilde{k}_{\perp}^2w_0^2}{4z_R}\right) \right. \\
\left.+\mathrm{e}^{-\frac{(\tau/2)^2}{4}(\omega+\tilde{k}_0)^2}\delta\left(\tilde{k}_z-\tilde{k}_0-\frac{\tilde{k}_{\perp}^2w_0^2}{4z_R}\right)\right].
\end{multline}
Notice that for long pulses, $\tau \gg 1$, the biggest contribution to the integral is provided by $\tilde{k}_0\approx\pm\omega$ and, hence when substituting $z_R$, the expression above can be approximated by 
\begin{multline}\label{a0}
\tilde{a}_0(\tilde{k},k)=-\frac{i\pi\sqrt{\pi}w_0^2\mathcal{E}_0}{\tilde{k}_0}\frac{\tau}{2}\mathrm{e}^{-\frac{\tilde{k}_{\perp}^2w_0^2}{4}} \delta\left(\tilde{k}_z-\tilde{k}_0+\frac{\tilde{k}_{\perp}^2}{2\tilde{k}_0}\right)\times \\
\left[ \mathrm{e}^{-\frac{(\tau/2)^2}{4}(\omega-\tilde{k}_0)^2} +\mathrm{e}^{-\frac{(\tau/2)^2}{4}(\omega+\tilde{k}_0)^2}\right],
\end{multline}
which coincides with the corresponding result found in \cite{Waters} for $\tilde{k}_{\perp}\ll\tilde{k}_0$. Moreover, in the limit $\tau\to\infty$ we use an exponential representation of the Dirac delta function, $\lim_{\epsilon\to0}\frac{1}{\sqrt{2\pi\epsilon}}\mathrm{e}^{-\frac{x^2}{2\epsilon}}=\delta(x)$, and obtain
\begin{multline}
\tilde{a}_0(\tilde{k},k)=-\frac{i2\pi^3w_0^2\mathcal{E}_0}{\tilde{k}_0}\mathrm{e}^{-\frac{\tilde{k}_{\perp}^2w_0^2}{4}} \delta\left(\tilde{k}_z-\tilde{k}_0+\frac{\tilde{k}_{\perp}^2}{2\tilde{k}_0}\right)\times \\
\left( \delta(\omega-\tilde{k}_0) +\delta(\omega+\tilde{k}_0)\right),
\end{multline}
which corresponds to the vector potential of a Gaussian beam.


\begin{thebibliography}{58}
\expandafter\ifx\csname
natexlab\endcsname\relax\def\natexlab#1{#1}\fi
\expandafter\ifx\csname bibnamefont\endcsname\relax
  \def\bibnamefont#1{#1}\fi
\expandafter\ifx\csname bibfnamefont\endcsname\relax
  \def\bibfnamefont#1{#1}\fi
\expandafter\ifx\csname citenamefont\endcsname\relax
  \def\citenamefont#1{#1}\fi
\expandafter\ifx\csname url\endcsname\relax
  \def\url#1{\texttt{#1}}\fi
\expandafter\ifx\csname urlprefix\endcsname\relax\def\urlprefix{URL
}\fi \providecommand{\bibinfo}[2]{#2}
\providecommand{\eprint}[2][]{\url{#2}}


\bibitem{BreitWheeler}
\bibinfo{author}{\bibfnamefont{G.}~\bibnamefont{Breit}} \bibnamefont{and}
\bibinfo{author}{\bibfnamefont{J.~A.}~\bibnamefont{Wheeler}},
\emph{\bibinfo{Title}{Collision of Two Light Quanta}},
\bibinfo{journal}{Phys. Rev.} \pmb{\bibinfo{volume}{46}},
\bibinfo{pages}{1087} (\bibinfo{year}{1934}).

\bibitem{Burke}
\bibinfo{author}{\bibfnamefont{D.~L.}~\bibnamefont{Burke}}\emph{\bibinfo{Title}{ et al.}},
\emph{\bibinfo{Title}{Positron production in multiphoton light-by-light scattering}},
\bibinfo{journal}{Phys. Rev. Lett.} \pmb{\bibinfo{volume}{79}},
\bibinfo{pages}{1626} (\bibinfo{year}{1997}).

\bibitem{Bamber}
\bibinfo{author}{\bibfnamefont{C.}~\bibnamefont{Bamber}}\emph{\bibinfo{Title}{ et al.}},
\emph{\bibinfo{Title}{Studies of nonlinear QED in collisions of 44.6 GeV electrons with intense laser pulses}},
\bibinfo{journal}{Phys. Rev. D} \pmb{\bibinfo{volume}{60}},
\bibinfo{pages}{092004} (\bibinfo{year}{1999}).

\bibitem{RitusReview}
\bibinfo{author}{\bibfnamefont{V.~I.}~\bibnamefont{Ritus}}, 
\emph{\bibinfo{Title}{Quantum effects of the interaction of elementary particles with an intense electromagnetic field}},
\bibinfo{journal}{J. Sov. Laser Res.} \pmb{\bibinfo{volume}{6}},
 \bibinfo{pages}{497} (\bibinfo{year}{1985}).
 
\bibitem{Reiss2009}
\bibinfo{author}{\bibfnamefont{H.~R.}~\bibnamefont{Reiss}},
\emph{\bibinfo{Title}{Special analytical properties of ultrastrong coherent fields}},
\bibinfo{journal}{Eur. Phys. J. D} \pmb{\bibinfo{volume}{55}},
\bibinfo{pages}{365} (\bibinfo{year}{2009}).

\bibitem{Hu2010}
\bibinfo{author}{\bibfnamefont{H.}~\bibnamefont{Hu}},\bibinfo{author} {\bibfnamefont{ C.}~\bibnamefont{M\"uller}}
\bibnamefont{and}
\bibinfo{author}{\bibfnamefont{C.~H.}~\bibnamefont{Keitel}},
\emph{\bibinfo{Title}{Complete QED theory of multiphoton trident pair production in strong laser fields}},
\bibinfo{journal}{Phys. Rev. Lett.} \pmb{\bibinfo{volume}{105}},
\bibinfo{pages}{080401} (\bibinfo{year}{2010}).

\bibitem{EhlotzkyReview}
\bibinfo{author}{\bibfnamefont{F.}~\bibnamefont{Ehlotzky}},
\bibinfo{author}{\bibfnamefont{K.}~\bibnamefont{Krajewska}} 
\bibnamefont{and} \bibinfo{author}{\bibfnamefont{J.~Z.}~\bibnamefont{Kami\'nski}},
\emph{\bibinfo{Title}{Fundamental processes of quantum
electrodynamics in laser fields of relativistic power}},
\bibinfo{journal}{Rep. Prog. Phys.} \pmb{\bibinfo{volume}{72}},
\bibinfo{pages}{046401} (\bibinfo{year}{2009}).

\bibitem{diPiazzaReview}
\bibinfo{author}{\bibfnamefont{A.}~\bibnamefont{Di Piazza}},
\bibinfo{author}{\bibfnamefont{C.}~\bibnamefont{M\"uller}}, \bibinfo{author}{\bibfnamefont{K.~Z.}~\bibnamefont{Hatsagortsyan}}
\bibnamefont{and} \bibinfo{author}{\bibfnamefont{C.~H.}~\bibnamefont{Keitel}},
\emph{\bibinfo{Title}{Extremely high-intensity laser interactions with fundamental quantum systems}},
\bibinfo{journal}{Rev. Mod. Phys.} \pmb{\bibinfo{volume}{84}},
\bibinfo{pages}{1177} (\bibinfo{year}{2012}).

\bibitem{Krajewska2012}
\bibinfo{author}{\bibfnamefont{K.}~\bibnamefont{Krajewska}} 
\bibnamefont{and} \bibinfo{author}{\bibfnamefont{J.~Z.}~\bibnamefont{Kami\'nski}},
\emph{\bibinfo{Title}{Breit-Wheeler process in intense short laser pulses}},
\bibinfo{journal}{Phys. Rev. A} \pmb{\bibinfo{volume}{86}},
\bibinfo{pages}{052104} (\bibinfo{year}{2012}).

\bibitem{Krajewska2014}
\bibinfo{author}{\bibfnamefont{K.}~\bibnamefont{Krajewska}} 
\bibnamefont{and} \bibinfo{author}{\bibfnamefont{J.~Z.}~\bibnamefont{Kami\'nski}},
\emph{\bibinfo{Title}{Coherent combs of antimatter from non-linear electron-positron-pair creation}},
\bibinfo{journal}{Phys. Rev. A} \pmb{\bibinfo{volume}{90}},
\bibinfo{pages}{052108} (\bibinfo{year}{2014}).

\bibitem{diPiazza}
\bibinfo{author}{\bibfnamefont{A.}~\bibnamefont{Di Piazza}},
\emph{\bibinfo{Title}{Nonlinear Breit-Wheeler Pair Production in a Tightly Focused Laser Beam}},
\bibinfo{journal}{Phys. Rev. Lett.} \pmb{\bibinfo{volume}{117}},
\bibinfo{pages}{213201} (\bibinfo{year}{2016}).

\bibitem{Meuren}
\bibinfo{author}{\bibfnamefont{S.}~\bibnamefont{Meuren}},
\bibinfo{author}{\bibfnamefont{C. H.}~\bibnamefont{Keitel}} \bibnamefont{and}
\bibinfo{author}{\bibfnamefont{A.}~\bibnamefont{Di Piazza}},
\emph{\bibinfo{Title}{Semiclassical picture for electron-positron photoproduction in strong laser fields}},
\bibinfo{journal}{Phys. Rev. D} \pmb{\bibinfo{volume}{93}},
\bibinfo{pages}{085028} (\bibinfo{year}{2016}).

\bibitem{Blackbourn}
\bibinfo{author}{\bibfnamefont{T.~G.}~\bibnamefont{Blackburn}} \bibnamefont{and}
\bibinfo{author}{\bibfnamefont{M.}~\bibnamefont{Marklund}},
\emph{\bibinfo{Title}{Nonlinear Breit-Wheeler pair creation with bremsstrahlung $\gamma$ rays}},
\bibinfo{journal}{Plasma Phys. Control. Fusion} \pmb{\bibinfo{volume}{60}},
\bibinfo{pages}{054009} (\bibinfo{year}{2018}).

\bibitem{Selym}
\bibinfo{author}{\bibfnamefont{S.}~\bibnamefont{Villalba-Ch\'avez}} 
\bibnamefont{and} \bibinfo{author}{\bibfnamefont{C.}~\bibnamefont{M\"uller}},
\emph{\bibinfo{Title}{Photo-production of scalar particles in the field of a circularly polarized laser beam}},
\bibinfo{journal}{Phys. Lett. B} \pmb{\bibinfo{volume}{718}},
\bibinfo{pages}{992} (\bibinfo{year}{2013}).

\bibitem{Jansen2013}
\bibinfo{author}{\bibfnamefont{M.~J.~A.}~\bibnamefont{Jansen}}, 
\bibnamefont{and} \bibinfo{author}{\bibfnamefont{C.}~\bibnamefont{M\"uller}},
\emph{\bibinfo{Title}{Strongly enhanced pair production in combined high- and low-frequency laser fields}},
\bibinfo{journal}{Phys. Rev. A} \pmb{\bibinfo{volume}{88}},
\bibinfo{pages}{052125} (\bibinfo{year}{2013}).

\bibitem{Heinzl}
\bibinfo{author}{\bibfnamefont{T.}~\bibnamefont{Heinzl}},
\bibinfo{author}{\bibfnamefont{A.}~\bibnamefont{Ilderton}} \bibnamefont{and}
\bibinfo{author}{\bibfnamefont{M.}~\bibnamefont{Marklund}},
\emph{\bibinfo{Title}{Finite size effects in stimulated laser pair production}},
\bibinfo{journal}{Phys. Lett. B} \pmb{\bibinfo{volume}{692}},
\bibinfo{pages}{250} (\bibinfo{year}{2010}).

\bibitem{Kaempfer2012}
\bibinfo{author}{\bibfnamefont{A.~I.}~\bibnamefont{Titov}},
\bibinfo{author}{\bibfnamefont{H.}~\bibnamefont{Takabe}},
\bibinfo{author}{\bibfnamefont{B.}~\bibnamefont{K\"ampfer}} \bibnamefont{and}
\bibinfo{author}{\bibfnamefont{A.}~\bibnamefont{Hosaka}},
\emph{\bibinfo{Title}{Enhanced subthreshold e+e- production in short laser pulses}},
\bibinfo{journal}{Phys. Rev. Lett.} \pmb{\bibinfo{volume}{108}},
\bibinfo{pages}{240406} (\bibinfo{year}{2012}).

\bibitem{Kaempfer2016}
\bibinfo{author}{\bibfnamefont{A.~I.}~\bibnamefont{Titov}},
\bibinfo{author}{\bibfnamefont{B.}~\bibnamefont{K\"ampfer}}, 
\bibinfo{author}{\bibfnamefont{A.}~\bibnamefont{Hosaka}}, \bibinfo{author}{\bibfnamefont{T.}~\bibnamefont{Nousch}} \bibnamefont{and}
\bibinfo{author}{\bibfnamefont{D.}~\bibnamefont{Seipt}},
\emph{\bibinfo{Title}{Determination of the carrier envelope phase for short, circularly polarized laser pulses}},
\bibinfo{journal}{Phys. Rev. D} \pmb{\bibinfo{volume}{93}},
\bibinfo{pages}{045010} (\bibinfo{year}{2016}).

\bibitem{Jansen2016}
\bibinfo{author}{\bibfnamefont{M.~J.~A.}~\bibnamefont{Jansen}},
\bibinfo{author}{\bibfnamefont{J.~Z.}~\bibnamefont{Kami\'nski}},
\bibinfo{author}{\bibfnamefont{K.}~\bibnamefont{Krajewska}} \bibnamefont{and}
\bibinfo{author}{\bibfnamefont{C.}~\bibnamefont{M\"uller}},
\emph{\bibinfo{Title}{Strong-field Breit-Wheeler pair production in short laser pulses: Relevance of spin effects}},
\bibinfo{journal}{Phys. Rev. D} \pmb{\bibinfo{volume}{94}},
\bibinfo{pages}{013010} (\bibinfo{year}{2016}).

\bibitem{Jansen2017}
\bibinfo{author}{\bibfnamefont{M.~J.~A.}~\bibnamefont{Jansen}} \bibnamefont{and}
\bibinfo{author}{\bibfnamefont{C.}~\bibnamefont{M\"uller}},
\emph{\bibinfo{Title}{Strong-field Breit–Wheeler pair production in two consecutive laser pulses with variable time delay}},
\bibinfo{journal}{Phys. Lett. B} \pmb{\bibinfo{volume}{766}},
\bibinfo{pages}{71} (\bibinfo{year}{2017}).

\bibitem{Grobe}
\bibinfo{author}{\bibfnamefont{Q.~Z.}~\bibnamefont{Lv}},
\bibinfo{author}{\bibfnamefont{S.}~\bibnamefont{Dong}}, \bibinfo{author}{\bibfnamefont{Y. T.}~\bibnamefont{Li}},
\bibinfo{author}{\bibfnamefont{Z. M.}~\bibnamefont{Sheng}}, \bibinfo{author}{\bibfnamefont{Q.}~\bibnamefont{Su}} 
\bibnamefont{and} \bibinfo{author}{\bibfnamefont{R.}~\bibnamefont{Grobe}},
\emph{\bibinfo{Title}{Role of the spatial
inhomogeneity on the laser-induced vacuum decay}},
\bibinfo{journal}{Phys. Rev. A} \pmb{\bibinfo{volume}{97}},
\bibinfo{pages}{022515} (\bibinfo{year}{2018}).

\bibitem{Kaempfer2018}
\bibinfo{author}{\bibfnamefont{A.~I.}~\bibnamefont{Titov}},
\bibinfo{author}{\bibfnamefont{H.}~\bibnamefont{Takabe}} \bibnamefont{and}
\bibinfo{author}{\bibfnamefont{B.}~\bibnamefont{K\"ampfer}},
\emph{\bibinfo{Title}{ Breit-Wheeler process in short laser double pulses}},
\bibinfo{journal}{Phys. Rev. D} \pmb{\bibinfo{volume}{98}},
\bibinfo{pages}{036022} (\bibinfo{year}{2018}).

\bibitem{CommentLinear}
\bibinfo{author}{\bibfnamefont{The term 'linear' refers to the linear dependence of the pair production rate on the applied photon beam intensity}}.

\bibitem{KingGies}
\bibinfo{author}{\bibfnamefont{B.}~\bibnamefont{King}},
\bibinfo{author}{\bibfnamefont{H.}~\bibnamefont{Gies}} \bibnamefont{and}
\bibinfo{author}{\bibfnamefont{A.}~\bibnamefont{Di Piazza}},
\emph{\bibinfo{Title}{Pair production in a plane wave by thermal background photons}},
\bibinfo{journal}{Phys. Rev. D} \pmb{\bibinfo{volume}{86}},
\bibinfo{pages}{125007} (\bibinfo{year}{2012}).

\bibitem{Pike}
\bibinfo{author}{\bibfnamefont{O.J.}~\bibnamefont{Pike}}\emph{\bibinfo{Title}{ et al.}},
\emph{\bibinfo{Title}{A photon-photon collider in a vacuum hohlraum}},
\bibinfo{journal}{Nature Photonics} \pmb{\bibinfo{volume}{8}},
\bibinfo{pages}{434} (\bibinfo{year}{2014}).

\bibitem{Drebot}
\bibinfo{author}{\bibfnamefont{I.}~\bibnamefont{Drebot}}\emph{\bibinfo{Title}{ et al.}},
\emph{\bibinfo{Title}{Matter from light-light scattering via Breit-Wheeler events produced by two interacting Coulomb sources}},
\bibinfo{journal}{Phys. Rev. Accel. Beams} \pmb{\bibinfo{volume}{20}},
\bibinfo{pages}{043402} (\bibinfo{year}{2017}).

\bibitem{Ribeyre}
\bibinfo{author}{\bibfnamefont{X.}~\bibnamefont{Ribeyre}}\emph{\bibinfo{Title}{ et al.}},
\emph{\bibinfo{Title}{Pair creation in collision of $\gamma$-ray beams produced with high-intensity lasers}},
\bibinfo{journal}{Phys. Rev. E} \pmb{\bibinfo{volume}{93}},
\bibinfo{pages}{013201} (\bibinfo{year}{2016}).

\bibitem{Yu}
\bibinfo{author}{\bibfnamefont{I.~J.}~\bibnamefont{Yu}}\emph{\bibinfo{Title}{ et al.}},
\emph{\bibinfo{Title}{Creation of Electron-Positron Pairs in Photon-Photon Collisions Driven by 10-PW Laser Pulses}},
\bibinfo{journal}{Phys. Rev. Lett.} \pmb{\bibinfo{volume}{122}},
\bibinfo{pages}{014802} (\bibinfo{year}{2019}).

\bibitem{Golub}
\bibinfo{author}{\bibfnamefont{A.}~\bibnamefont{Golub}},
\bibinfo{author}{\bibfnamefont{R.}~\bibnamefont{Egger}}, 
\bibinfo{author}{\bibfnamefont{C.}~\bibnamefont{M\"uller}}\bibnamefont{and}
\bibinfo{author}{\bibfnamefont{S.}~\bibnamefont{Villalba-Ch\'avez}},
\emph{\bibinfo{Title}{Dimensionality-Driven Photoproduction of Massive Dirac Pairs near Threshold in Gapped Graphene Monolayers}},
\bibinfo{journal}{Phys. Rev. Lett.} \pmb{\bibinfo{volume}{124}},
\bibinfo{pages}{110403} (\bibinfo{year}{2020}).

\bibitem{Piran}
\bibinfo{author}{\bibfnamefont{T.}~\bibnamefont{Piran}},
\emph{\bibinfo{Title}{The physics of gamma-ray bursts}},
\bibinfo{journal}{Rev. Mod. Phys.} \pmb{\bibinfo{volume}{76}},
\bibinfo{pages}{1143} (\bibinfo{year}{2005}).

\bibitem{Ruffini}
\bibinfo{author}{\bibfnamefont{R.}~\bibnamefont{Ruffini}},
\bibinfo{author}{\bibfnamefont{V.~G.}~\bibnamefont{Vereshchagin}}\bibnamefont{and}
\bibinfo{author}{\bibfnamefont{S.~S.}~\bibnamefont{Xue}},
\emph{\bibinfo{Title}{Electron–positron pairs in physics and astrophysics: From heavy nuclei to black holes}},
\bibinfo{journal}{Phys. Rep.} \pmb{\bibinfo{volume}{487}},
\bibinfo{pages}{1} (\bibinfo{year}{2010}).

\bibitem{Leemanns2019}
\bibinfo{author}{\bibfnamefont{A.~J.}~\bibnamefont{Gonsalves}}\emph{\bibinfo{Title}{ et al.}},
\emph{\bibinfo{Title}{Petawatt Laser Guiding and Electron Beam Acceleration to 8 GeV
in a Laser-Heated Capillary Discharge Waveguide}},
\bibinfo{journal}{Phys. Rev. Lett.} \pmb{\bibinfo{volume}{122}},
\bibinfo{pages}{084801} (\bibinfo{year}{2019}).

\bibitem{Karsch}
\bibinfo{author}{\bibfnamefont{G.}~\bibnamefont{G\"otzfried}}\emph{\bibinfo{Title}{ et al.}},
\emph{\bibinfo{Title}{Physics of nanocoulomb-class electron beams in laser-plasma wakefields}},
\bibinfo{journal}{arXiv:2004.10310v1}.

\bibitem{HiBEF}
\emph{\bibinfo{Title}{http://www.hibef.eu/}}.

\bibitem{Chen}
\bibinfo{author}{\bibfnamefont{H.}~\bibnamefont{Chen}}\emph{\bibinfo{Title}{ et al.}},
\emph{\bibinfo{Title}{Relativistic Positron Creation Using Ultraintense Short Pulse Lasers}},
\bibinfo{journal}{Phys. Rev. Lett.} \pmb{\bibinfo{volume}{102}},
\bibinfo{pages}{105001} (\bibinfo{year}{2009}).

\bibitem{Sarri}
\bibinfo{author}{\bibfnamefont{G.}~\bibnamefont{Sarri}}\emph{\bibinfo{Title}{ et al.}},
\emph{\bibinfo{Title}{Table-Top Laser-Based Source of Femtosecond, Collimated, Ultrarelativistic Positron Beams}},
\bibinfo{journal}{Phys. Rev. Lett.} \pmb{\bibinfo{volume}{110}},
\bibinfo{pages}{255002} (\bibinfo{year}{2013}).

\bibitem{Fradkin}
\bibinfo{author}{\bibfnamefont{E. S.}~\bibnamefont{Fradkin}},
\bibinfo{author}{\bibfnamefont{D. M.}~\bibnamefont{Gitman}} \bibnamefont{and}
\bibinfo{author}{\bibfnamefont{S. V.}~\bibnamefont{Shvartsman}},
\emph{\bibinfo{Title}{Quantum Electrodynamics with Unstable Vacuum}},
\bibinfo{journal}{Springer-Verlag, Berlin Heidelberg} (\bibinfo{year}{1961}).

\bibitem{Greiner}
\bibinfo{author}{\bibfnamefont{W.}~\bibnamefont{Greiner}} \bibnamefont{and}
\bibinfo{author}{\bibfnamefont{J.}~\bibnamefont{Reinhardt}},
\emph{\bibinfo{Title}{Quantum Electrodynamics}},
\bibinfo{journal}{Springer-Verlag, Berlin Heidelberg} (\bibinfo{year}{2003}).

\bibitem{Waters}
\bibinfo{author}{\bibfnamefont{W. J.}~\bibnamefont{Waters}} \bibnamefont{and}
\bibinfo{author}{\bibfnamefont{B.}~\bibnamefont{King}},
\emph{\bibinfo{Title}{On beam models and their paraxial approximation}},
\bibinfo{journal}{Laser Phys.} \pmb{\bibinfo{volume}{28}},
\bibinfo{pages}{015003} (\bibinfo{year}{2018}).

\bibitem{Blinne}
\bibinfo{author}{\bibfnamefont{A.}~\bibnamefont{Blinne}}\emph{\bibinfo{Title}{ et al.}},
\emph{\bibinfo{Title}{Photon-Photon Scattering at the High-Intensity Frontier: Paraxial Beams}},
\bibinfo{journal}{J. Phys.: Conf. Ser.} \pmb{\bibinfo{volume}{1206}},
\bibinfo{pages}{012016} (\bibinfo{year}{2019}).

\bibitem{Stearns}
\bibinfo{author}{\bibfnamefont{M.}~\bibnamefont{Stearns}},
\emph{\bibinfo{Title}{Mean Square Angles of Bremsstrahlung and Pair Production}},
\bibinfo{journal}{Phys. Rev. } \pmb{\bibinfo{volume}{76}},
 \bibinfo{pages}{836} (\bibinfo{year}{1949}).

\bibitem{Tsai}
\bibinfo{author}{\bibfnamefont{Y.-S.}~\bibnamefont{Tsai}},
\emph{\bibinfo{Title}{Pair production and bremsstrahlung of charged leptons}},
\bibinfo{journal}{Rev. Mod. Phys.} \pmb{\bibinfo{volume}{46}},
 \bibinfo{pages}{815} (\bibinfo{year}{1974}).

\bibitem{Leemans2014}
\bibinfo{author}{\bibfnamefont{W.~P.}~\bibnamefont{Leemans}}\emph{\bibinfo{Title}{ et al.}},
\emph{\bibinfo{Title}{Multi-GeV Electron Beams from Capillary-Discharge-Guided Subpetawatt Laser Pulses in the Self-Trapping Regime}},
\bibinfo{journal}{Phys. Rev. Lett.} \pmb{\bibinfo{volume}{113}},
\bibinfo{pages}{245002} (\bibinfo{year}{2014}).

\bibitem{LCLS}
\emph{\bibinfo{Title}{https://lcls.slac.stanford.edu/parameters}}.
\end{thebibliography}
\end{document}